\documentclass[12pt]{article}
\usepackage{pstricks}
\usepackage{color}
\usepackage{cite}
\usepackage{array}
\usepackage{epsfig}
\usepackage{amssymb}
\usepackage{graphics,graphpap}
\usepackage{amssymb}
\usepackage{amsmath}
\usepackage{slashed}
\usepackage{dsfont}
\usepackage{a4wide}
 \usepackage{array,multirow,graphicx}
 \usepackage{float}

\def\eps{\varepsilon}
\def\epe{\varepsilon'/\varepsilon}

\newcommand{\gev}{\, {\rm GeV}}
\newcommand{\mev}{\, {\rm MeV}}

\newcommand{\mt}{m_{\rm t}}
\newcommand{\mtb}{\overline{m}_{\rm t}}

\newcommand{\be}{\begin{equation}}
\newcommand{\ee}{\end{equation}}
\newcommand{\bea}{\begin{eqnarray}}
\newcommand{\eea}{\end{eqnarray}}

\newcommand{\bi}{\begin{itemize}}
\newcommand{\ei}{\end{itemize}}
\newcommand{\ord}{{\cal O}}

\newcommand{\vcb}{|V_{cb}|}
\newcommand{\vtd}{|V_{td}|}
\newcommand{\vub}{|V_{ub}|}
\newcommand{\vts}{|V_{ts}|}
\newcommand{\vus}{|V_{us}|}

\def\kpn{K^+\rightarrow\pi^+\nu\bar\nu}

\def\klpn{K_{L}\rightarrow\pi^0\nu\bar\nu}

\usepackage{graphicx}

\medskipamount % besser als explizite Angabe in pt; etwas Platz nach neuem Absatz

 \def\s#1{\setbox0=\hbox{$#1$}%
   \rlap{\ifdim\wd0>.7em\kern.22\wd0\else\kern.1\wd0\fi /}#1}


\begin{document}

%%%%%%%%%% Title page
\begin{titlepage}
\begin{flushright}
\begin{tabular}{l}
TTP18-044 \\
{AJB-18-10}
\end{tabular}
\end{flushright}
\vskip1.2cm
\begin{center}
{\Large \bf \boldmath
Emerging $\Delta M_{d}$-Anomaly from Tree-Level \vspace{2mm} \\Determinations of $|V_{cb}|$  
and the Angle $\gamma$}
\vskip1.0cm
{\bf \large
Monika Blanke$^{a,b}$ and Andrzej J. Buras$^{c,d}$}
\vskip0.3cm
$^a$ {Institut f\"ur Kernphysik, Karlsruhe Institute of Technology,
  Hermann-von-Helmholtz-Platz 1,
  D-76344 Eggenstein-Leopoldshafen, Germany}\vspace{1mm}\\
$^b$ {Institut f\"ur Theoretische Teilchenphysik,
  Karlsruhe Institute of Technology, Engesserstra\ss e 7,
  D-76128 Karlsruhe, Germany}\vspace{1mm}\\
$^c$ TUM-IAS, Lichtenbergstr. 2a, D-85748 Garching, Germany\vspace{1mm}\\
$^d$Physik Department, TUM, D-85748 Garching, Germany\\

\vskip0.51cm

%{\em Version of \today}

\vskip0.35cm

{\large\bf Abstract\\[10pt]} \parbox[t]{\textwidth}{
We point out that the recently increased value of the angle $\gamma$ 
in the Unitarity Triangle (UT), determined in tree-level decays to be $\gamma=(74.0^{+5.0}_{-5.8})^\circ$ by the LHCb collaboration, combined with the most recent value of $|V_{cb}|$ implies an enhancement of $\Delta M_{d}$ over the data in the ballpark of $30\%$. This is {larger by roughly a factor of two} than the enhancement of $\Delta M_{s}$ that is independent of $\gamma$.
This disparity of enhancements is  problematic for models with Constrained Minimal Flavour Violation (CMFV) and also for $U(2)^3$ models. In view of the prospects of measuring $\gamma$ with the precision of $\pm 1^\circ$ by Belle II and LHCb in the coming years, we 
propose to use the angles $\gamma$ and $\beta$ together with $|V_{cb}|$ and $|V_{us}|$ 
as the fundamental parameters of the CKM matrix until $|V_{ub}|$ from tree-level 
decays will be known precisely. 
Displaying $\Delta M_{s,d}$ as functions of $\gamma$ clearly demonstrates the tension between the value of $\gamma$ from tree-level decays, free from new physics (NP) contributions, and $\Delta M_{s,d}$ calculated in CMFV and  $U(2)^3$  models and thus exhibits the presence of NP contributions to  $\Delta M_{s,d}$ beyond these frameworks. We calculate the values of $|V_{ub}|$ and $|V_{td}|$ as functions of $\gamma$ and $|V_{cb}|$ and 
discuss the implications of our results for $\varepsilon_K$ and rare $K$ and $B$ decays. 
We also briefly discuss  a future strategy in which $\beta$, possibly affected by NP, is replaced by $|V_{ub}|$.
}

\end{center}
\end{titlepage}

\setcounter{footnote}{0}

\newpage

\section{Introduction}
\label{sec:1}

The $\Delta F=2$ transitions in the 
down-quark sector, that is $B^0_{s,d}-\bar B^0_{s,d}$ and 
$K^0-\bar K^0$ mixings, have been vital in constraining the {Standard Model (SM)} and in the search for new physics (NP) for several decades\cite{Isidori:2010kg,Buras:2013ooa}. However, theoretical uncertainties related to 
the hadronic matrix elements entering these transitions and their large sensitivity to the CKM parameters made clear cut conclusions about the presence of NP impossible. As we demonstrate in this paper, this could change
in the near future.

Among the most important flavour observables we have {at} our disposal are
\be\label{great5}
\Delta M_s,\quad \Delta M_d, \quad S_{\psi K_S},\quad S_{\psi \phi}, \quad \varepsilon_K
\ee
with $\Delta M_{s,d}$ being the mass differences in 
$B^0_{s,d}-\bar B^0_{s,d}$ mixings and $S_{\psi K_S}$ and $S_{\psi \phi}$ 
the corresponding mixing induced CP-asymmetries.
$\varepsilon_K$ describes the magnitude of indirect CP-violation in $K^0-\bar K^0$ mixing. $\Delta M_{s,d}$ and  $\varepsilon_K$ 
are already known experimentally with impressive precision. The asymmetries  $S_{\psi K_S}$ and $S_{\psi \phi}$ are less precisely  measured but have the advantage of being subject to only very small hadronic uncertainties. 

On the other hand the CKM parameters of particular interest are
\be\label{CKMpar}
\vus,\qquad\vcb,\qquad \vub,\qquad \gamma,\qquad \beta,
\ee
with the first three being the moduli of the most intensively studied elements of the 
CKM matrix, and $\gamma$ and $\beta$ being two angles in the Unitarity Triangle (UT).
The angle $\gamma$ is to an excellent approximation equal to the sole complex
phase in the standard parametrization of the CKM matrix.

Now, as elaborated in \cite{Buras:2002yj}, there are many ways to construct the rescaled 
UT. They all involve only two inputs, but as quantified in the latter paper, 
some pairs are particularly suited for the determination of
the apex ($\bar\rho,\bar\eta$) of this triangle, as only moderate precision 
on them is required to obtain a satisfactory determination of $\bar\rho$ 
and $\bar\eta$. The clear winners from this study are the pairs
\be
(\beta,\gamma), \qquad (R_b,\gamma),
\ee
with $R_b$ being the length of one side in the UT related to the ratio $\vub/\vcb$.

Ideally, one would like to use the second pair which allows to construct the 
so-called {reference unitarity triangle} (RUT) \cite{Goto:1995hj} 
that is supposed to be free of NP contributions. Unfortunately, the persistent discrepancy between inclusive and exclusive determinations of $\vub$ from tree-level decays precludes a satisfactory determination of the RUT at present. 

On the other hand the tree-level determination of the angle $\gamma$ has  significantly been
improved in the last years by various measurements of the LHCb collaboration, with
the latest average being \cite{Kenzie:2319289}\footnote{The HFLAV average $\gamma = (73.5^{+4.2}_{-5.1})^\circ$ \cite{Amhis:2016xyh} does not include the latest LHCb result.} 
\be\label{LHCbgamma}
\gamma=(74.0^{+5.0}_{-5.8})^\circ,\,.
\ee
Moreover, the prospects of LHCb and Belle II \cite{Kou:2018nap,Krishnan:2018gdn} to decrease the error down to $\pm 1^\circ$ are promising. In view of this situation and significant recent progress in the determination of $\vcb$, giving \cite{Gambino:2016jkc}
\be\label{Gambino}
\vcb=(42.0\pm0.6)\times 10^{-3}\,,
\ee
we will choose as the four fundamental CKM parameters
\be\label{CKMparbest}
\vus,\qquad\vcb,\qquad \gamma,\qquad \beta\,.
\ee

Within the SM and CMFV models \cite{Buras:2000dm,Buras:2003jf,Blanke:2006ig}, the hadronic uncertainties in   $\Delta M_{s,d}$ reside within a good approximation in the parameters
\be\label{hpar}
 F_{B_s}\sqrt{\hat B_{B_s}},\quad  F_{B_d} \sqrt{\hat B_{B_d}}.
\ee
Fortunately, during the last years their uncertainties decreased significantly.
In particular, an impressive progress has been made by the Fermilab Lattice and MILC Collaborations (Fermilab-MILC) 
 that find \cite{Bazavov:2016nty}
\be\label{Kronfeld}
 F_{B_s}\sqrt{\hat B_{B_s}}=(274.6\pm8.8)\mev,\qquad  F_{B_d} \sqrt{\hat B_{B_d}}=
(227.7\pm 9.8)\mev \,,
\ee
with uncertainties of $3\%$ and $4\%$, respectively. An even higher 
precision is achieved for the ratio 
\be\label{xi}
\xi=\frac{F_{B_s}\sqrt{\hat B_{B_s}}}{F_{B_d}\sqrt{\hat B_{B_d}}}=1.206\pm0.019\,.
\ee

Based on the results in (\ref{Kronfeld}) and (\ref{xi}) we have performed in 
\cite{Blanke:2016bhf} a detailed analysis of $\Delta F=2$ processes in CMFV 
models, finding a significant tension between $\Delta M_{s,d}$ 
and $\varepsilon_K$ in these models with the pattern of the tension 
strongly dependent on the value of $\vcb$. Moreover, {constructing the universal unitarity triangle (UUT) \cite{Buras:2000dm} via $R_t$ and $\beta$} we could predict, independently of $\vcb$, the value of $\gamma$ to be
\be
\gamma=(63.0\pm 2.1)^\circ\,,
\ee
significantly below the value in (\ref{LHCbgamma}). {This number has not changed with respect to our 2016 analysis, and now displays a $1.8\sigma$ tension with the improved tree level measurement in \eqref{LHCbgamma}.\footnote{Using instead of \eqref{xi} the very recent RBC-UKQCD result $\xi = 1.1853\pm 0.0054^{+0.0116}_{-0.0156}$ \cite{Boyle:2018knm}, one obtains $\gamma = (60.7 \pm 1.5)^\circ$, increasing the tension to
 $2.2 \sigma$.}} As we have discussed 
in \cite{Blanke:2016bhf} this problem arises not only in the SM and more generally in CMFV models but also in minimally broken $U(2)^3$ models, where NP contributions in the $B_d$ and $B_s$ systems are universal and hence cancel in the ratio.

 As the present paper deals again  with the tensions between $\Delta F=2$ 
observables in CMFV models, it is mandatory for us to state what is new in our
paper:
\begin{itemize}
\item
In \cite{Blanke:2016bhf}, we have considered two strategies. One in which $\varepsilon_K$ has been used to determine $\vcb$, implying a value consistent with the inclusive determination as well as  $\Delta M_{s,d}$ values well above the data. In the second strategy, $\vcb$ has been determined from $\Delta M_s$ resulting in a low value of $\vcb$ consistent with the exclusive determination at that time. The predicted $\varepsilon_K$ then turned 
out to be well below its experimental value. The recent improvements in the determinations of 
$\vcb$ \cite{Bigi:2017njr,Bigi:2017jbd} disfavours the second strategy and also the recent claim in \cite{Bailey:2018feb} that there is a $4\sigma$ anomaly in $\varepsilon_K$.
\item
More importantly, in view of the improved value of $\gamma$, 
we decided to use it as an input in the present analysis, instead of the usual determination of the UT in CMFV models through $S_{\psi K_S}~(\beta)$ and the side $R_t$ of the UT determined from the ratio $\Delta M_{d}/\Delta M_{s}$ and $\xi$ in (\ref{xi}).
\item
The most recent discussions, see in particular \cite{DiLuzio:2017fdq,DiLuzio:2018wch}, dealt exclusively with the implications of the enhanced value of $\Delta M_s$ and 
not $\Delta M_d$, for which  in addition to the increased value of $\vcb$ also 
the increased value of $\gamma$ matters.
\end{itemize}

In the context of the second item we remark that the $(R_t,\beta)$ strategy 
for the determination of the UT has been found in \cite{Buras:2002yj} to be  less
 powerful than the  $(\beta,\gamma)$ strategy used here. Moreover, as NP 
now is expected in $\Delta M_{s,d}$, it appears as a better strategy to replace 
their ratio by the angle $\gamma$ and instead treat $\Delta M_{s,d}$ as outputs being  functions of $\gamma$, $\beta$ and $\vcb$.

{One could wonder why the emerging $\Delta M_d$ anomaly pointed out by us has not been noticed in the global fits performed by the CKMfitter and UTfit collaborations. In our
view such global fits, involving simultaneously many quantities, are 
likely to miss NP effects present in only a subset of observables, in particular when the significance has not reached the discovery level. We are optimistic  that the findings of this paper pointing towards NP in the $B_d$ system will motivate both theorists and experimentalists to intensify the
search for NP in $b\to d$ transitions, after the last five years being dominated 
by the study of $b\to s$ and $b\to c$ transitions.}

Our paper is organized as follows. In Section~\ref{sec:2} we present the determination of the UT and of the CKM matrix using the $(\beta,\gamma)$ strategy. 
In Section~\ref{sec:3}  we evaluate  $\Delta M_{d}$ and $\Delta M_{s}$ as functions of $\gamma$,
finding their values to disagree with the data. The new result
relative to \cite{Blanke:2016bhf} and other recent papers  
\cite{DiLuzio:2017fdq,DiLuzio:2018wch} is the 
disagreement of $\Delta M_{d}$ and the ratio  $\Delta M_{d}/\Delta M_{s}$ with the data, a direct 
consequence of the increased value of $\gamma$. On the other hand $\varepsilon_K$ agrees well with the data. We therefore provide the SM predictions for the branching ratios of $\kpn$ and $\klpn$ for different values of $\gamma$, $\beta$ and $\vcb$. In Section~\ref{sec:4} we have a look at the $(R_b,\gamma)$ strategy, which could become favourable in the next decade, once the tree-level determination of $|V_{ub}|$ is settled.
In Section~\ref{sec:5} we briefly investigate  what kind of NP could be responsible for the
$\Delta M_{s,d}$ anomalies found in Section~\ref{sec:3} and what are the implications for NP in $\Delta F =1$ transitions.  We conclude in Section~\ref{sec:6}.

\section{Deriving the UT and the CKM matrix}\label{sec:2}

Our determination of the UT and of  the CKM matrix proceeds in two steps:

{\bf Step 1:}

We use as input parameters 
\be\label{beta}
\beta=(21.85\pm0.67)^\circ, \qquad \gamma=(74.0^{+5.0}_{-5.8})^\circ,
\ee
with $\beta$ 
obtained from 
\be\label{SpKs}
S_{\psi K_S}=\sin 2\beta =0.691 \pm 0.017 \,.
\ee

\begin{figure}
\centering
\includegraphics[width = 0.55\textwidth]{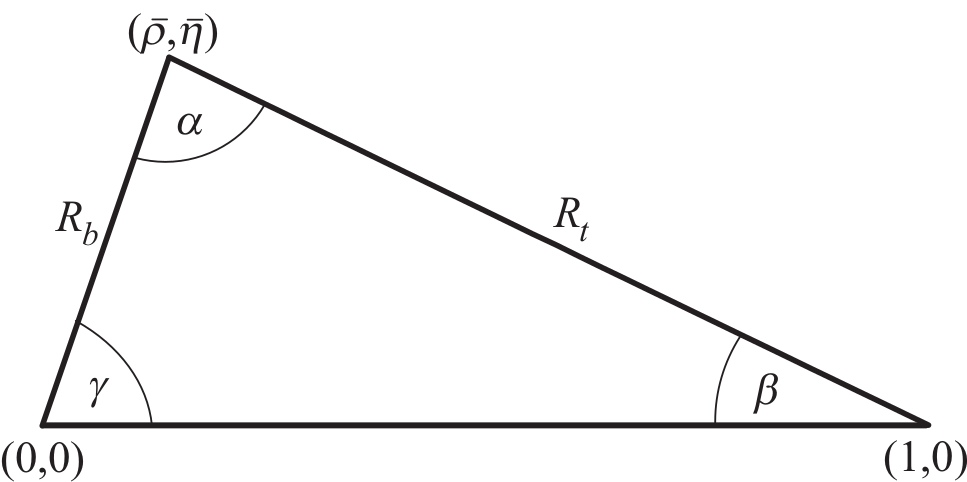}
 \caption{\it The Unitarity Triangle. }\label{UUTa}
\end{figure}

This allows us to determine the two sides $R_b$ and $R_t$ of the UT 
shown in Fig.~\ref{UUTa}, that are given in terms of $\beta$ and $\gamma$ as follows  \cite{Buras:2002yj}
\begin{equation}\label{2.96}
R_b=\frac{\sin(\beta)}{\sin(\gamma+\beta)}=0.374\pm 0.012\,,\qquad
R_t=
\frac{\sin(\gamma)}{\sin(\gamma+\beta)}=0.964\pm 0.035\,.
\end{equation}

The angles $\beta$ and $\gamma$ of the unitarity triangle are directly related
 to the complex phases of the CKM-elements $V_{td}$ and
$V_{ub}$, respectively, through
\begin{equation}\label{e417}
V_{td}=|V_{td}|e^{-i\beta},\quad V_{ub}=|V_{ub}|e^{-i\gamma}.
\end{equation}

{\bf Step 2:}

Including $\lambda \equiv \vus$ and $\vcb$ as the remaining input parameters we determine 
$\vtd$ and $\vts$ through
\be\label{simple}
\vtd=\vus\vcb R_t\,,\qquad \vts=\eta_R\vcb~
\ee
with
\be
\eta_R=1 -\vus\xi\sqrt{\frac{\Delta M_d}{\Delta M_s}}\sqrt{\frac{m_{B_s}}{m_{B_d}}}\cos\beta+\frac{\lambda^2}{2}+\ord(\lambda^4) = {0.9825}\,,
\ee
where we have used $\beta$ in (\ref{beta}). $\Delta M_{d,s}$ are taken from 
 experiment as given in Table~\ref{tab:input} but using our $\vcb$ and $\gamma$ dependent 
values would change the result by less than $1\%$.

Finally, we find
\begin{equation}\label{2.94}
\vub=\lambda \vcb \frac{R_b}{1-\frac{\lambda^2}{2}}\,.
\end{equation}

\begin{figure}
\centering{\includegraphics[width=.8\textwidth]{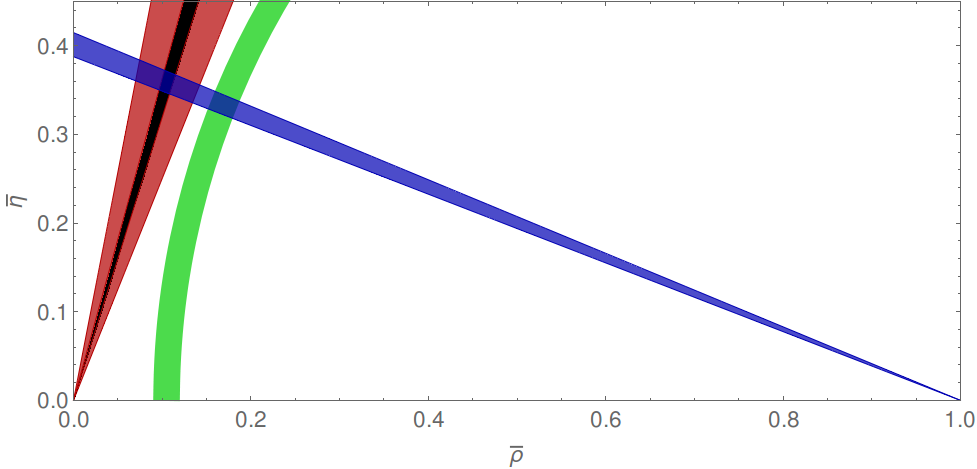}}
\caption{\it \label{fig:UTconst} Constraints on the UT from the angles $\gamma$ (red) and $\beta$ from $S_{\psi K_S}$ (blue), and $R_t$ from $\Delta M_d/\Delta M_s$ (green).}
\end{figure}

In Fig.~\ref{fig:UTconst} we show  the constraints on the UT from the tree-level 
measurement of $\gamma$, from  $\beta$ extracted from $S_{\psi K_S}$, and $R_t$ from $\Delta M_d/\Delta M_s$. The advantage of the $(\gamma,\beta)$ strategy 
over the $(R_t,\beta)$ strategy is not seen yet because of a significant error 
in $\gamma$. With the future uncertainty on $\gamma$ of $\pm 1^\circ$ represented by the black area, 
the power of the $(\gamma,\beta)$ strategy in determining the UT is clearly 
visible. However, already now we observe that the apex of the UT 
obtained from the $(\gamma,\beta)$ strategy disagrees with the one from the 
$(R_t,\beta)$ one. {This tension indicates the presence of some NP contributions.}

\begin{figure}
\includegraphics[width=.47\textwidth]{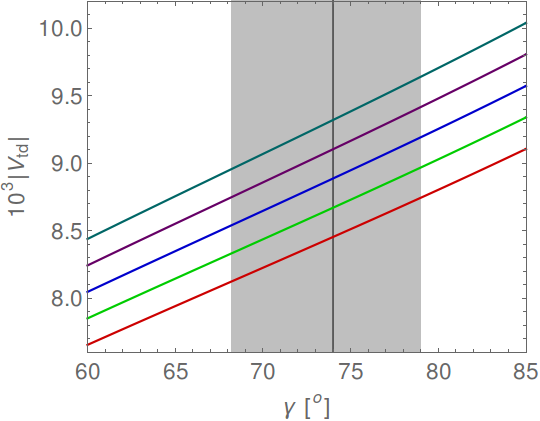}
\hfill
\includegraphics[width=.47\textwidth]{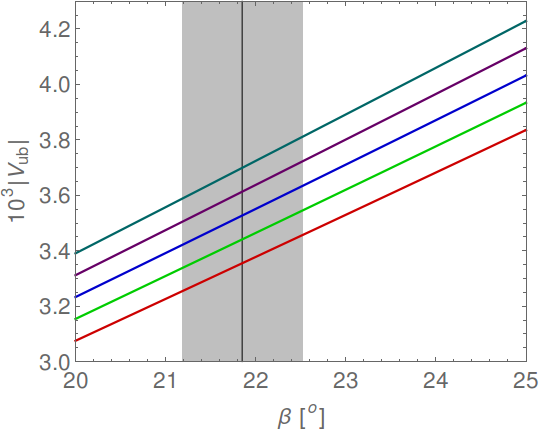}
\caption{\label{fig:VtdVub}\it Left: $|V_{td}|$ as function of $\gamma$, for different values of $|V_{cb}|$. Right: $|V_{ub}|$ as function of $\beta$, 
for different values of $|V_{cb}|$. The colours correspond to: $|V_{cb}|=39\cdot 10^{-3}$ (red, bottom), $40\cdot 10^{-3}$ (green), $41\cdot 10^{-3}$ (blue), $42\cdot 10^{-3}$ (purple), $43\cdot 10^{-3}$ (turquoise, top).}
\end{figure}

In Fig.~\ref{fig:VtdVub} we show $\vtd$ as a function of $\gamma$ and $\vub$ 
as a function of $\beta$ for different values of $\vcb$. 
The dependences of  $\vtd$ on $\beta$ and of $\vub$ on $\gamma$ are very small.
These plots will allow to monitor the values of $\vtd$ and $\vub$ that 
enter various observables as the uncertainties of $\gamma$, $\beta$ and $\vcb$
will shrink with time.

\section{Calculating observables}\label{sec:3}

 For the mass differences in the $B^0_{s,d}-\bar B^0_{s,d}$ systems we have 
{the very accurate expressions} \cite{Blanke:2016bhf} 
\bea
\label{DMD}
\Delta M_d&=&
0.5055/{\rm ps}\cdot\left[ 
\frac{\sqrt{\hat B_{B_d}}F_{B_d}}{227.7\mev}\right]^2
\left[\frac{S(v)}{2.322}\right]
\left[\frac{\vtd}{8.00\cdot10^{-3}} \right]^2 
\left[\frac{\eta_B}{0.5521}\right]\,, \\
\label{DMS}
\Delta M_{s}&=&
17.757/{\rm ps}\cdot\left[ 
\frac{\sqrt{\hat B_{B_s}}F_{B_s}}{274.6\mev}\right]^2
\left[\frac{S(v)}{2.322}\right]
\left[\frac{\vts}{0.0390} \right]^2
\left[\frac{\eta_B}{0.5521}\right] \,.
\eea
Here  $S(v)$ is the box-function in CMFV models with $v$ denoting 
parameters of a given model including $x_t=m_t^2/M_W^2$.
The value $2.322$ in the normalization of $S(v)$ is its SM value for 
$m_t(m_t)=163.5\gev$ {obtained from}
\be\label{S0}
S_0(x_t)  = \frac{4x_t - 11 x_t^2 + x_t^3}{4(1-x_t)^2}-\frac{3 x_t^2\log x_t}{2
(1-x_t)^3}= 2.322 \left[\frac{\mtb(\mt)}{163.5\gev}\right]^{1.52}\,,
\ee
and $\eta_B$ is the perturbative QCD correction \cite{Buras:1990fn}. Our input parameters, equal 
to the ones used in \cite{Bazavov:2016nty}, are collected in Table~\ref{tab:input}. 
{We find
\be
(\Delta M_d)_\text{SM} =  (0.648 \pm 0.077)\,\text{ps}^{-1}\,,\qquad
(\Delta M_s)_\text{SM} = (19.8 \pm 1.4)\,\text{ps}^{-1}\,,
\ee
which differ from the experimental values by $1.9\sigma$ and $1.4\sigma$, respectively. As the correlation matrix of the relevant lattice parameters entering these predictions is unknown to us, we do not attempt to derive a global significance for the anomaly in $B_{d,s}-\bar B_{d,s}$ mixing. }

\begin{table}[!tb]
\center{\begin{tabular}{|l|l|}
\hline
$m_{B_s} = 5366.8(2)\mev$\hfill\cite{Agashe:2014kda}	&  $m_{B_d}=5279.58(17)\mev$\hfill\cite{Agashe:2014kda}\\
 $\Delta M_s = 17.757(21) \,\text{ps}^{-1}$\hfill\cite{Amhis:2016xyh}	&  $\Delta M_d = 0.5055(20) \,\text{ps}^{-1}$\hfill\cite{Amhis:2016xyh}\\ 
$S_{\psi K_S}= 0.691(17)$\hfill\cite{Amhis:2016xyh}
		&  $S_{\psi\phi}= 0.015(35)$\hfill \cite{Amhis:2016xyh}\\
	$|V_{us}|=0.2253(8)$\hfill\cite{Agashe:2014kda} &
 $|\eps_K|= 2.228(11)\cdot 10^{-3}$\hfill\cite{Agashe:2014kda}\\
$F_{B_s}$ = $228.6(3.8)\mev$ \hfill \cite{Rosner:2015wva} & $F_{B_d}$ = $193.6(4.2)\mev$ \hfill \cite{Rosner:2015wva}  \\
$m_t(m_t)=163.53(85)\gev$ & $S_0(x_t)=2.322(18)$ \\
$\eta_{cc}=1.87(76)$\hfill\cite{Brod:2011ty} & $\eta_{ct}= 0.496(47)$\hfill\cite{Brod:2010mj}\\
$\eta_{tt}=0.5765(65)$\hfill\cite{Buras:1990fn}	&
$\eta_B=0.55(1)$\hfill\cite{Buras:1990fn,Urban:1997gw}\\
$\tau_{B_s}= 1.510(5)\,\text{ps}$\hfill\cite{Amhis:2016xyh} &  $\Delta\Gamma_s/\Gamma_s=0.124(9)$\hfill\cite{Amhis:2016xyh} \\
$\tau_{B_d}= 1.520(4)\,\text{ps}$\hfill\cite{Amhis:2016xyh} &  $\kappa_\varepsilon = 0.94(2)$\hfill \cite{Buras:2010pza}
\\	       
\hline
\end{tabular}  }
\caption {\textit{Values of the experimental and theoretical
    quantities used as input parameters. For future 
updates see PDG \cite{Agashe:2014kda}  and HFLAV \cite{Amhis:2016xyh}. 
}}
\label{tab:input}
\end{table}

Now, the overall factors in (\ref{DMD}) and (\ref{DMS}) are the central experimental values, and in CMFV models $S(v)$ is bounded from below by its SM value \cite{Blanke:2006yh}
\be\label{BBBOUND}
S(v)\ge S_0(x_t)= 2.322\,.
\ee
{Consequently, with the values of $\vtd$ found in the previous section, 
that are significantly larger than its nominal value in (\ref{DMD}), it is evident that CMFV models have difficulties in describing the data for $\Delta M_d$. 
In addition, with the value of $\vcb$ in (\ref{Gambino}) also $\vts$ is significantly larger than its nominal value in (\ref{DMS}). Therefore $\Delta M_s$ in CMFV models is enhanced over its experimental value as already pointed out in 
 \cite{Bazavov:2016nty,Blanke:2016bhf} and recently analysed in
\cite{DiLuzio:2017fdq,DiLuzio:2018wch}. Yet the latter enhancement is not as large 
as for $\Delta M_d$ because $\Delta M_s$ does not depend on $\gamma$.}

\begin{figure}
\includegraphics[width=.47\textwidth]{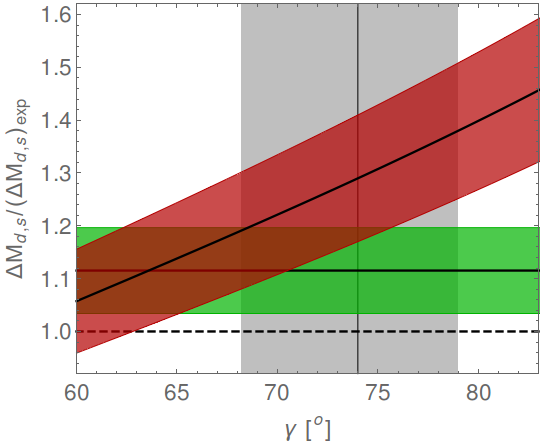}
\hfill
\includegraphics[width=.47\textwidth]{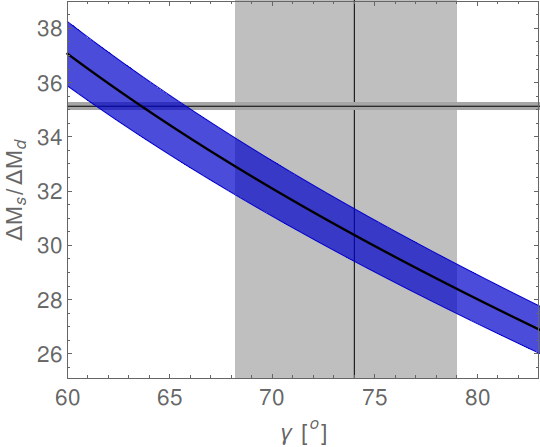}
\caption{\it\label{fig:ga-DMds} Left: $\Delta M_d$ (red) and $\Delta M_s$ (green) as functions of $\gamma$, normalised to their experimental values. The $1\sigma$-band includes all other uncertainties. Right: $\Delta M_s/\Delta M_d$ as function of $\gamma$.}
\end{figure}

In Fig.~\ref{fig:ga-DMds} we show in the left panel $\Delta M_d$ and $\Delta M_s$ normalized to {their} experimental values. Evidently, for 
central values of all parameters, $\Delta M_d$ differs  by roughly $30\%$ from the data while 
in the case of $\Delta M_s$ the corresponding difference amounts only to $12\%$.
But the uncertainties in other parameters like $\vcb$ and the hadronic parameters in (\ref{Kronfeld}) are {still} significant. However, we expect that in the coming years these uncertainties will {significantly} be reduced. 

{In the right panel of Fig.~\ref{fig:ga-DMds} we show the ratio $\Delta M_s/\Delta M_d$ as a function of $\gamma$. The dependence on $\vcb$ cancels in this 
ratio and the error on $\xi$ in (\ref{xi}) is much smaller than the errors in 
(\ref{Kronfeld}). Consequently the disagreement of the ratio in question with the data, shown as a horizontal line at $35.1$, is 
clearly visible and expresses the problem of CMFV models and those based on the
  $U(2)^3$  symmetry.}

\begin{figure}
\centering{\includegraphics[width=.47\textwidth]{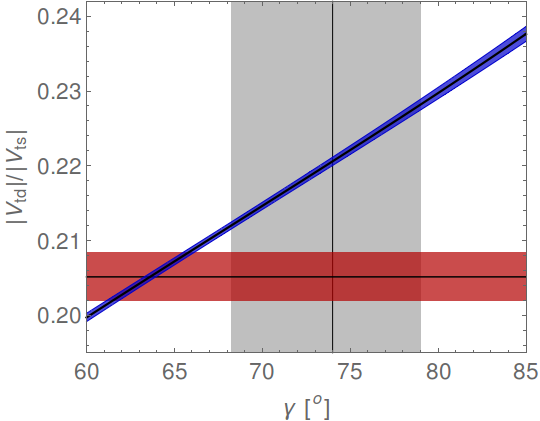}}
\caption{\it\label{fig:Vtds} The ratio $|V_{td}|/|V_{ts}|$ as function of the angle $\gamma$. In blue we show the prediction from eq.\, \eqref{DMSDMD0}, to be compared with the result of \eqref{DMSDMD} displayed in red.}
\end{figure}

Of interest is also the ratio
\be\label{DMSDMD0}
\frac{\vtd}{\vts}=\xi\sqrt{\frac{m_{B_s}}{m_{B_d}}}\sqrt{\frac{\Delta M_d}{\Delta M_s}}
\ee
with $\Delta M_{s,d}$  predicted here to be compared with
\be\label{DMSDMD}
\frac{\vtd}{\vts}=\xi\sqrt{\frac{m_{B_s}}{m_{B_d}}}\sqrt{\frac{(\Delta M_d)_{\rm exp}}{(\Delta M_s)_{\rm exp}}}= {0.2052\pm 0.0033}\,,
\ee
with $\Delta M_{s,d}$ taken from experiment. {In CMFV models and those
with  $U(2)^3$  symmetry this ratio depends only on the angle $\gamma$. We 
show this in Fig. \ref{fig:Vtds}. A significant enhancement of $|V_{td}|/|V_{ts}|$ over the value in (\ref{DMSDMD}) 
is observed.}

{As far as $\varepsilon_K$ is concerned, using the standard expression as given e.\,g.\ in 
\cite{Buras:2013ooa} and all input parameters collected in Table~\ref{tab:input}, we find the SM value of $\varepsilon_K$ to be fully consistent with the data:
\be
|\varepsilon_K|_{\rm SM}= (2.26\pm 0.27) \cdot 10^{-3}\,,
\ee
with higher values for the remaining CMFV models due to the bound in (\ref{BBBOUND}).}

Despite the fact that the SM fails to describe the data for $\Delta M_{s,d}$, having determined the CKM parameters,
the agreement of the SM with the experimental value for $\varepsilon_K$ invites us to calculate the branching ratios for $\kpn$ and $\klpn$ in the SM. This is 
of interest in view of the NA62 and KOTO experiments that should provide 
results for these decays in the coming years.
{Using 
the parametric formulae of \cite{Buras:2015qea} we find the central values of
$\mathcal{B}(\kpn)$ and $\mathcal{B}(\klpn)$
given in table \ref{tab:kpinunu} for different values of $\gamma$, $\beta$ and $\vcb$.}

\begin{table}
\begin{minipage}{.49\textwidth}
\center{$10^{11}\cdot\mathcal{B}(K^+\to\pi^+\nu\bar\nu)$

\begin{tabular}{|c|ccccc|}
\hline
$\gamma [^\circ]$& \multicolumn{5}{c|}{$10^3\cdot |V_{cb}|$} \\
 & 39 & 40 & 41 & 42 & 43 \\\hline
64 & 6.7 & 7.2 & 7.8 & 8.3 & 8.9 \\
66 & 6.9 & 7.4 & 7.9 & 8.5 & 9.1 \\
68 & 7.1 & 7.6 & 8.1 & 8.7 & 9.3 \\
70 & 7.2 & 7.7 & 8.3 & 8.9 & 9.5 \\
72 & 7.4 & 7.9 & 8.5 & 9.1 & 9.7 \\
74 & 7.5 & 8.1 & 8.6 & 9.2 & 9.9 \\
76 & 7.7 & 8.2 & 8.8 & 9.4 & 10.0\\
78 & 7.8 & 8.4 & 9.0 & 9.6 & 10.3\\	       
\hline
\end{tabular}  }
\end{minipage}
\hfill
\begin{minipage}{.49\textwidth}
\center{$10^{11}\cdot\mathcal{B}(K_L\to\pi^0\nu\bar\nu)$

\begin{tabular}{|c|c|ccccc|}
\hline
$\beta [^\circ]$ & $\gamma [^\circ]$& \multicolumn{5}{c|}{$10^3\cdot |V_{cb}|$} \\
 && 39 & 40 & 41 & 42 & 43 \\\hline
\multirow{4}{*}{$ 21.85$} 
& 65 & 2.1 & 2.3 & 2.5 & 2.8 & 3.1 \\
& 69 & 2.2 & 2.4 & 2.7 & 3.0 & 3.2 \\
& 73 & 2.3 & 2.6 & 2.8 & 3.1 & 3.4 \\
& 77 & 2.4 & 2.7 & 3.0 & 3.3 & 3.6 \\
 \hline
 \multirow{4}{*}{$ 24.0$}
& 65 & 2.5 & 2.7 & 3.0 & 3.3 & 3.7 \\
& 69 & 2.6 & 2.9 & 3.2 & 3.5 & 3.9 \\
& 73 & 2.8 & 3.1 & 3.4 & 3.8 & 4.1 \\
& 77 & 3.0 & 3.3 & 3.6 & 4.0 & 4.4 \\
 \hline
\end{tabular}  }
\end{minipage}
\caption {\textit{Left: Central values for the branching ratio $\mathcal{B}(K^+\to\pi^+\nu\bar\nu)$ for various values of $\gamma$ and $|V_{cb}|$. The angle $\beta$ is fixed to $\beta =21.85^\circ$ determined from $S_{\psi K_S}$.
Right: Central values for the branching ratio $ \mathcal{B}(K_L\to\pi^0\nu\bar\nu)$ for various values of $\beta$, $\gamma$ and $|V_{cb}|$.
}}
\label{tab:kpinunu}
\end{table}

\boldmath
\section{$(R_b,\gamma)$ strategy}\label{sec:4}
\unboldmath

It is likely that in the next decade the $(\beta,\gamma)$ strategy will be 
replaced by the $(R_b,\gamma)$ strategy. This could turn out to be even necessary if the 
value of $\vub$ determined from tree-level processes turned out to be
very different from the one determined in the previous section. Therefore 
for completeness we want to give the relevant formulae for this strategy.

Knowing $\vub$ determined in tree-level decays, one finds $R_b$ using
\begin{equation}\label{2.94a}
R_b =\left(1-\frac{\lambda^2}{2}\right)\frac{1}{\lambda}
\left| \frac{V_{ub}}{V_{cb}} \right|\,.
\end{equation}
Together with $\gamma$, this result allows to determine $R_t$ and $\beta$ 
by means of 
\be\label{VTDG}
R_t=\sqrt{1+R_b^2-2 R_b\cos\gamma},\qquad
\cot\beta=\frac{1-R_b\cos\gamma}{R_b\sin\gamma}\,,
\ee
so that the RUT is completely fixed.

If the resulting value of $\beta$ differs from the one in (\ref{beta}), then
the expression in (\ref{SpKs}) will have to be replaced by 
\begin{equation}
S_{\psi K_S} = \sin(2\beta+2\varphi_{\rm new}) =0.691 \pm 0.017 \,,
\label{U21}
\end{equation}
with $\varphi_{\rm new}$ being a new CP-violating phase. For instance for $\vub=4.0\cdot 10^{-3}$ we find 
\be
\varphi_{\rm new} \simeq -2.2^\circ\,.
\ee

\section{Going beyond CMFV}\label{sec:5}

Our analysis signals the violation  of flavour universality in 
the function $S(v)$, characteristic for  CMFV models. It hints for the presence of new sources of flavour and CP-violation and/or new operators contributing to $\Delta F=2$ transitions beyond the SM $(V-A)\otimes (V-A)$ ones.\footnote{In a more general formulation of MFV new operators could be present \cite{D'Ambrosio:2002ex}.} 
For simplicity we restrict first
 our discussion of NP scenarios to the ones in which only SM operators 
are present.

A fully general and very convenient solution in this case is just to consider instead of the  flavour universal function $S(v)$ three functions
\be\label{BSM}
S_i=S_0(x_t)-\Delta S_i e^{i\delta_i}  \qquad  (i=K,s,d) \,,
\ee
with $\Delta S_i$ being real and positive definite quantities, and the minus 
sign required to suppress $\Delta M_{d,s}$ below their SM values. It is evident that with two free parameters in each meson system it is always possible to obtain an agreement with the data on $\Delta F=2$ observables.  Our analysis indicates the following pattern of these
parameters:
\begin{itemize}
\item
A clear breakdown of the universality of $S(v)$  with 
\be
 \Delta S_s < \Delta S_d\,.
\ee
\item
The new phases
\be
\delta_s\approx \delta_d\approx 0
\ee
in order not to spoil the good agreement of the SM with the experimental values 
of $S_{\psi K_S}$ and $S_{\psi \phi}$. 
\item{In the case of $K^0-\bar K^0$ mixing, the good agreement of $\varepsilon_K$ with its measured value implies a small imaginary part of the NP contribution. This can either be achieved by 
a small value of $\Delta S_K$, or by an appropriately chosen value of 
the new phase $\delta_K$.}
\end{itemize}

Note that the fate of $\delta_d$ and to a lesser extend of $\delta_K$ will
depend on the future value of $\vub$ as remarked in connection with (\ref{U21}).

\begin{figure}
\centering{\includegraphics[width=.47\textwidth]{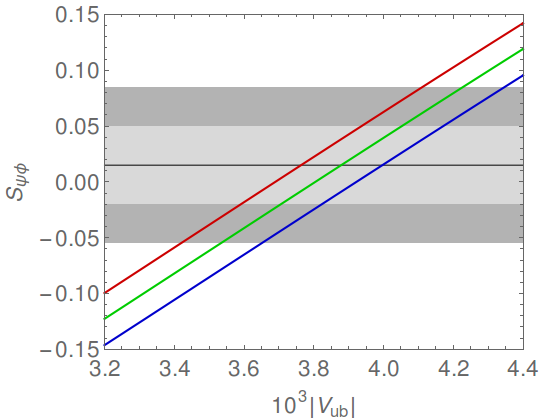}}
\caption{\it\label{fig:VubSpsiphi} Correlation between $|V_{ub}|$ and $S_{\psi\phi}$ in $U(2)^3$ models, for three different values of $S_{\psi K_S}$: $0.674$ (red), $0.691$ (green), $0.708$ (blue). The experimental $1\sigma$ and $2\sigma$ regions for $S_{\psi\phi}$ are shown by the grey bands.}
\end{figure}

This pattern cannot be explained in models with a minimally broken $U(2)^3$ flavour symmetry {\cite{Kagan:2009bn,Barbieri:2011ci,Barbieri:2012uh}} in which the equality
$\Delta S_s =\Delta S_d$ is predicted, although the near equality of 
$\delta_s$ and $\delta_d$ is a property of these models. This could change 
if for instance $\vub$ was found significantly different from the 
value followed from our strategy, {as shown in Fig.~\ref{fig:VubSpsiphi}}. But as these models fail anyway we will 
not consider them further.

The simplest models beyond the CMFV and $U(2)^3$  frameworks one could consider
are models with tree-level  $Z^\prime$ and $Z$ exchanges. While in 
\cite{Buras:2012jb,Buras:2015jaq,Endo:2016tnu,Bobeth:2017xry} general studies of such scenarios have been considered, specific examples are models with vector-like quarks  \cite{Bobeth:2016llm} and 331 models \cite{Buras:2016dxz}.  These models have sufficient numbers of parameters to obtain an agreement 
with the data for $\Delta F=2$ processes. This is explicitly shown for the case of 331 models in \cite{Buras:2016dxz}.

The minus sign in  (\ref{BSM}) has been introduced by us by hand. Strictly 
speaking, as already discussed in the context of $\Delta M_s$ in \cite{DiLuzio:2017fdq}, in the presence of only left-handed currents the minus sign in (\ref{BSM}) truly requires the NP phases to be $\pi+\delta_{d,s}$. 
{Following the reasoning in \cite{Blanke:2009pq}, this implies the CP-violating phases in the corresponding $\Delta F =1 $ $b\to d,s$ transitions to be close to $\pi/2$, i.\,e.\ maximal. We hence conclude that, within models with only left-handed currents, the observed suppression of $\Delta M_{d}$ and to a lesser extent $\Delta M_s$ implies significant deviations from the SM in CP-asymmetries of radiative and rare $b\to d$ and $b\to s$ decays. As quantitative predictions for these observables are model-dependent, we leave their thorough analysis for future work.}

In the presence of both left-
and right-handed couplings, on the other hand, the suppression of $\Delta M_d$ is much easier to
achieve without introducing large CP-violating phases. 
In this context probably 
most interesting are models in which the SMEFT operator $\mathcal{O}_{Hd}$ involving right-handed flavour violating couplings to down-quarks is generated at the NP scale. 
As demonstrated in \cite{Bobeth:2017xry}, the renormalisation group evolution 
to low-energy scales involving also left-handed currents present already within the SM  generates left-right $\Delta F=2$ operators representing FCNCs mediated by the $Z$ boson. At NLO this effect has also been discussed in \cite{Endo:2016tnu}.
An explicit realization of such a NP scenario is provided by
models with vector-like quarks  with an additional $U(1)$ gauge symmetry so that both tree-level $Z$ and $Z^\prime$ exchanges are present, and in some models of this type also
box diagram contributions with vector-like quarks, Higgs and other scalar and pseudoscalar exchanges are important \cite{Bobeth:2016llm}. The test 
of these scenarios is then mainly  offered through the correlations of 
$\Delta M_{d,s}$ with $\Delta F=1$ processes, that is rare $K$ or $B_{s,d}$ decays, the ratio $\epe$ and other observables. This is evident from the analyses in 
\cite{Bobeth:2016llm,Bobeth:2017xry} and once the data on $\gamma$, $\vcb$ and 
$\vub$ improve, could be an arena for further investigation of the implications 
of the $\Delta M_d$ anomaly pointed out here.

\section{Summary}\label{sec:6}

The main message of our paper is the emerging $\Delta M_d$ anomaly which is significantly larger than the $\Delta M_s$ one discussed in \cite{Blanke:2016bhf,DiLuzio:2017fdq,DiLuzio:2018wch}. Its fate will depend strongly on the improved values of $\gamma$ and 
$\vcb$ from tree-level decays and, to a lesser extent, on $\vub$, which is more 
relevant for the prediction of $\sin 2\beta$ in the SM. This anomaly, if confirmed,
will have implications for observables sensitive to $b\to d$ transitions 
like $b\to d \ell^+\ell^-$ and $b\to d \nu\bar\nu$ which will be explored 
by Belle II. It will open a new oasis of NP, analogous to the one related to the recent anomalies in $b\to s \ell^+\ell^-$ and their implications for 
 $b\to s \nu\bar\nu$ transitions. Depending on the NP flavour structure, it could also have implications
for $\kpn$ and $\klpn$.

\subsection*{Acknowledgements}

The research of AJB was fully supported by the DFG
cluster of excellence ``Origin and Structure of the Universe''.

\bibliographystyle{JHEP}
\bibliography{allrefs}
\end{document}